\documentclass[11pt,a4paper]{article}
\usepackage[english]{babel}

\usepackage[utf8]{inputenc}
\usepackage[T1]{fontenc}

\usepackage[a4paper, top=1in, bottom=1in, left=1.25in, right=1.25in]{geometry}

\usepackage[tagged, highstructure]{accessibility}
\usepackage{graphicx}
\usepackage{authblk}
\usepackage{amsmath,amssymb,amsthm}
\usepackage{url}
\usepackage{pifont}

\let\origref\ref
\def\ref#1{\textnormal{\origref{#1}}}

\title{DAO-enabled decentralized physical AI:\\A new paradigm for human-machine collaboration}

\author[1]{Mark C. Ballandies$^*$, 0000-0001-9350-9823}
\author[2]{Florian Spychiger, 0000-0001-6602-8450}
\author[3,4]{Uwe Serdült, 0000-0002-2383-3158}
\author[1]{Claudio J. Tessone, 0000-0001-7733-6221}

\affil[1]{Blockchain and Distributed Ledger Technologies, University of Zurich, Zurich, Switzerland}

\affil[2]{School of Management and Law, Institute for Organizational Viability, Zurich University of Applied Sciences, Winterthur, Switzerland}

\affil[3]{Center for Democracy Studies Aarau (ZDA), University of Zurich, Aarau, Switzerland}

\affil[4]{College of Information Science and Engineering, Ritsumeikan University, Osaka, Japan}

\date{\today} 

\begin{document}
\maketitle

\begin{abstract}
We propose DAO-enabled decentralized physical AI (DePAI), a democratic architecture for coordinating humans and autonomous machines in the operation and governance of physical–digital systems. We (1) synthesize foundations in blockchains, decentralized autonomous organizations (DAOs), and cryptoeconomics; (2) connect DAO design with digital-democracy research on deliberation and voting, showing how each can advance the other; (3) position DAO-governed decentralized physical infrastructure networks (DePIN) within a vertically integrated stack that links energy and sensing to connectivity, storage/compute, models, and robots; (4) show how these elements specify workflows that couple machine execution with human oversight, enabling enhanced self-organization of techno-socio-economic systems, which we call DePAI; and (5) analyze risks, including security, centralization, incentive failure, legal exposure, and the crowding-out of intrinsic motivation, and argue for value-sensitive design and continuously adaptive governance. DePAI offers a path to scalable, resilient self-organization that integrates physical infrastructure, AI, and community ownership under transparent rules, on-chain incentives, and permissionless participation, aiming to preserve human autonomy.

\footnotetext[1]{$^*$ Corresponding author: Author A, School of Computing, University of Leeds, Leeds, UK, E-mail: authorA@leeds.ac.uk}

\end{abstract}

\section{Introduction}
Distributed Ledger Technologies (DLTs), notably blockchain, represent a foundational shift in socio‑techno-economic coordination. Specifically, token incentives function as feedback signals that facilitate self‑organization and enable the efficient crowd-funding of capital expenditures; decentralized governance via co‑ownership and co‑creation lowers the operational expenditures of managing large‑scale complex systems; and distributed ledger properties such as transparency, immutability, and automation facilitate the creation of novel value‑aligned systems.

In particular, the merging of the previous concepts gave rise to a new and innovative organizational form referred to as Decentralized Autonomous Organization (DAO). DAOs embody democratic ideals in a permissionless, trustless environment by combining token incentives with shared governance and ledger properties to build resilient and efficient global organizations. 
For instance, MakerDAO secures over \$6.87 billion in collateral; BanklessDAO motivates more than $2500$ contributors to collaborate on a regular basis; Helium spans continents with its crowd‑built IoT network;  and Bitcoin's open consensus protocol can ward off attacks by state actors. Beyond these use cases, DAOs promise a new paradigm of human–machine collaboration, an emergent theme explored in this chapter.

To this end, we begin with essential background on blockchain, DAOs and cryptoeconomics (Section \ref{sec:background}), then illustrate the synergies between digital democracy research and DAO design (Section \ref{sec:daos_and_democracy}), and show how DAOs can facilitate this human–machine collaboration (Section \ref{sec:dao_human_machine}). We then situate this possibility in the context of key risks (Section \ref{sec:risks}) and conclude the work with an outlook on DAO-enabled human-machine collaboration (Section \ref{sec:conclusion}).

\section{Background: Blockchain, DAOs and Cryptoeconomics}
\label{sec:background}

Decentralized ledgers, token incentives, and novel governance mechanisms have converged to form a new class of socio‑techno-economic systems.  In this section, we first review the
underlying technology of blockchains, then explore how tokenized communities organize
themselves as Decentralized Autonomous Organizations (DAOs), and finally, we introduce cryptoeconomics, the interdisciplinary research field that unifies these layers.

\subsection{Blockchain}
Decentralized ledgers, exemplified by blockchain, function as databases in which information, typically transaction data, is stored in discrete blocks that are cryptographically linked together in a chain. Each new entry references earlier entries, creating an ordered transaction history.
Blockchain’s key innovation lies in how information is added to the database. By employing novel consensus mechanisms such as Proof-of-Work or Proof-of-Stake, data can be appended without trusted intermediaries while satisfying two essential properties: It ensures that only formally correct information by authorized parties is added, and guarantees that once information is recorded, it either cannot be changed or can only be altered transparently through predefined rules.
For instance, in Proof-of-Work as introduced by Satoshi Nakamato for Bitcoin \cite{nakamoto2008bitcoin}, network operators compete to solve a cryptographic challenge requiring computational effort and consequently time. The greater the computing power operators contribute, the higher their chances of winning, generating the required proof, appending new information in the form of a block to the blockchain, and receiving a predefined reward. The proofs are designed to depend on the entire linked block history and are also stored within a newly produced block itself; thus, altering an older block would change the required proof of this block but also all subsequent proofs of the following blocks as they depend on each other via the linked list. Thus, the chain of blocks with valid proofs is longer in the chain where the past transaction has not been modified than in the chain where it has been, which facilitates the core idea of the Proof-of-Work consensus mechanism: Honest network participants always accept the longest chain of blocks as the valid blockchain and ignore shorter ones. This longer chain represents the chain where the most computational work has been put in. Assuming a majority in computational power of honest nodes, these nodes will produce the longest blockchain.
Upon successfully appending information, operators receive newly created token units as compensation for their computational work. These tokens are necessary for network users to pay for adding their transactions to the blockchain. This creates a microeconomy, connecting those providing blockchain infrastructure and earning tokens for their work with those who require the very same tokens to utilize the blockchain. Consequently, third-party markets emerge where these tokens can be traded, granting them monetary value and thus making them an incentive for operators to join the network in the first place and facilitate its security via the consensus \cite{ballandies2022decrypting}.

This creates a scenario in which game-theoretic reasoning illustrates that network operators are incentivized to act honestly: Altering the blockchain requires an attacker to redo the proof-of-Work not only for the block they wish to change but also for every subsequent block, because each block is cryptographically linked. To succeed, the attacker must redo all past blocks and additionally outpace the rest of the network in creating new blocks, which effectively requires them to control more than half of the network's computational power, a feat demanding in the case of Bitcoin immense investments in hardware and electricity. Even if such resources were available, the potential gain remains limited. While an attack might enable double-spending (allowing a bitcoin token to be spent twice) or similar manipulations, it would severely undermine trust in the Bitcoin network, causing the currency’s value, including any bitcoins held by the attacker, to collapse. Rationally, an attacker would find it more profitable to allocate its computational power toward earning tokens legitimately instead. Also, because all data and interactions with the blockchain are transparent, honest network participants can detect such an attack, allowing the issue to be resolved off-chain through a fork \cite{hsieh2025impact} which means that the community decides not to accept to use the attacked blockchain and create a new instance (e.g. as done in "The DAO" hack, see Section \ref{sec:dao_innovations}).
Therefore, the combination of high resource costs, limited potential gains, off-chain mitigation strategies, and the risk of devaluing one's own assets makes these attacks economically irrational, as demonstrated in Bitcoin network which has not been successfully attacked in such a way since its inception in 2009, despite it becoming the 6th most valuable asset by market cap\footnote{https://companiesmarketcap.com/assets-by-market-cap/, last accessed: 2025-07-24}.

In the first generation of blockchains, led by Bitcoin, the transactions recorded in the distributed ledger primarily involved token or monetary transfers (e.g., sending token units from party A to party B). The second generation, introduced with Ethereum, expanded this functionality by enabling the inclusion of (Turing-complete) code on the blockchain which is referred to as smart contracts. This transformed the blockchain into a state machine, effectively turning it into a global computer. With each transaction appended, the state is updated.
Through this mechanism, users can deploy rules to the blockchain whose execution is enforced by the decentralized consensus network, which ensures they cannot be stopped or altered. 

The ability to store immutable, self-executing code on-chain paved the way for decentralized applications that leverage blockchain properties and are built directly on top of it.
For instance, Decentralized Finance applications let users swap tokens ('currencies') algorithmically on decentralized exchanges, removing the need for a central intermediary. Smart contracts also allow anyone to create these tokens, offering a powerful means to incentivize and coordinate large groups toward a shared goal \cite{dapp2021finance,ballandies2021finance}. In particular, tokens can act as information signals that help complex systems self-organize through real-time feedback.
Technically, a token is merely an on-chain object, fields track balances and methods govern transfers, so writing the code is straightforward and thus system designers can create these incentives without the need to deploy and maintain their own blockchain. This ease has driven a surge in token deployments spanning many designs. Some resemble fiat currencies, while others differ markedly by, for instance, having a fixed supply that limits inflation. Tokens can also be non-transferable (e.g., reputation badges) or non-fungible, where each unit is unique \cite{ballandies20251}. Also, they can be combined to span complex multi-token systems \cite{dapp2021finance,kalabic2023burn}. 
Further, utilizing smart contracts, rules for human and machine interactions can be implemented on-chain. In particular, novel forms of governance have been deployed on-chain ranging from liquid democracy, over participatory budgeting to square-root voting, optimistic governance and Futarchy, as illustrated in greater detail in Section \ref{sec:dao_innovations}. 
Also, by affording values such as transparency, immutability or permissionlessness in technological systems, blockchain can enable the value-sensitive design of these very systems \cite{ietto2023blockchain,ballandies2024advancing}. These characteristics position blockchain-based systems, particularly DAOs, as fertile platforms for democratic innovation.

\subsection{Decentralized Autonomous Organization (DAO)}

The merging of token incentives, decentralized governance and the properties of the underlying blockchain gave rise to a new and innovative organizational form referred to as Decentralized Autonomous Organization~(DAO).  These DAOs can be defined as i) decentralized, characterized by a broad community of interacting peers; ii) organized, possessing mechanisms for coordination and decision-making; and iii) autonomous, enabling permissionless actions of community members \cite{ballandies2024daos}. This three-fold definition maps DAOs to their three core dimensions: Community, Governance and Execution \cite{makode2025taxonomy,lustenberger2024designing}. 
In particular, by merging the concepts of a community with that of an organization, DAOs can adapt efficiently to changing environments in a bottom-up and self-organized manner \cite{ballandies2024daos,lustenberger2024designing,lustenberger2024mastering}.
The community consists of individuals having realtionships with each other and where every individual has personal interstests and goals. The organization is the coordinated collaboration of these individuals. Within the organization, these individuals align towards a shared vision. Too much community can result in chaos and too much organization can result in restrictive structures harming community. However, if well designed, "a sense of community brings trust and wellbeing which enable better organization. And effective organization can enable value flows that sustain community" \cite{vanderMolenOspina2023} resulting in a resilient and efficient organization. In particular, DAOs are complex systems in which the decentralized community can facilitate collective intelligence, the coordination and decision-making mechanism in its governance can enable efficient digital democracy and the autonomous execution of tasks of community members can facilitate resilient adaptation to changing environments \cite{ballandies2024daos}.  

In practice, DAOs facilitate collective intelligence by being open, permissionless, and transparent. Anyone can join or exit freely; all information on governance and execution is public; and participants enjoy pseudonymous privacy and unrestricted expression.
This stems from hosting discussions on permissionless platforms (email lists, real-world conferences, Discord, Reddit, X, Telegram, etc.), where contributions under privacy‑preserving pseudonyms or non-identity revealing access mechanisms incur minimal cost in expression and encourage full participation.
The digital democracy in DAOs relies on Improvement Proposals - structured, peer‑reviewed documents open to anyone, followed by formal votes weighted by tokens, delegated stakes, or hash power of contributed computational resources.
Adaptation emerges in DAOs from real‑time feedback loops: token rewards incentivize bug reports, content creation, and consensus participation; an open‑source codebase ensures permissionless autonomy; and, without a decisive central entity, potentially even without a legal wrapper, members have permissionless autonomy to act. 
Together, these design choices can enable DAOs to transform distributed expertise into efficient governance and resilient adaptation.

When designing a DAO, the DAO design canvas can be utilized which separates the design process into three parts i) Purpose, ii) Use Case and iii) Legislation \cite{lustenberger2024mastering}.
Current research on DAOs is mainly focused on \cite{lustenberger2025dao}: i) analyzing governance mechanisms in DAOs, ii) developing tools and frameworks for DAOs, iii) assessing the value proposition of DAOs, and iv) exploring legal and regulatory dimensions in DAOs. 
What often is overlooked in research is the adaptation capability of DAOs to changing environments and the deliberation mechanisms applied within these organizations \cite{ballandies2025bitcoin}.

\subsection{Cryptoeconomics}
\label{sec:cryptoeconomics}
Cryptoeconomics centers around the design and analysis of token incentives, decentralized governance mechanisms, and blockchain protocol configurations. It is an emerging interdisciplinary discipline \cite{voshmgir2020foundations,ballandies2022fundamentals} that examines blockchains as complex socio-techno-economic systems. It highlights how the emergent behavior of these systems results not only from their technical configurations but also significantly from user interactions and embedded economic incentives \cite{spychiger2023incentive}. In particular, two blockchains with identical technological setups can exhibit varying degrees of immutability, a phenomenon explored in Section \ref{sec:dao_innovations}.  The following frameworks and taxonomies offer a foundation for deeper analysis and research in this area \cite{tasca2019taxonomy,voshmgir2020foundations,spychiger2021unveiling,ballandies2022decrypting,ballandies2022fundamentals}.

\section{DAOs and Democracy}
\label{sec:daos_and_democracy}

DAOs represent an innovation within digital democracy contributing novel insights and findings to the construction of democratic systems. On the other hand, by being embedded in this larger field of research, findings from democracy research can be applied to the design and instantiation of DAOs that are currently overlooked by practitioners. In the following, both aspects will be explored in greater detail.
\subsection{DAO innovations}
\label{sec:dao_innovations}


One of the fundamental design decisions that went into Bitcoin, arguably the first DAO \cite{hsieh2018bitcoin,ballandies2025bitcoin}, was that of permissionlessness. Contributors to the network do not have to ask permission for joining and contributing. Anybody can connect their hardware and start contributing to the network's security and be rewarded for it. In particular, no one can exclude members from the organization in the name of Bitcoin. Adapting to changing environments in a decentralized manner relies on individual actors responding quickly to local conditions \cite{dapp2021finance}. Without a governing body capable of censoring, directing, or manipulating such responses, one can argue, based on Bitcoin’s success, that this decentralized adaptability was instrumental in its growth and resilience~\cite{buterin2013bootstrapping,ballandies2025bitcoin}.

This permissionless nature has also given rise to a variety of emergent governance patterns, one is "forking"~\cite{hsieh2025impact}. In case a governance decision is not accepted by a part of the community, this part can decide to split from the rest of the community and continue with their own rules that match their requirements. This can also be initiated by a minority within the community. This was observed famously in "The DAO" hack in 2016. A hacker exploited a bug in a smart contract and extracted a large sum of tokens from it. The amount was so large that it posed a serious threat to the overall Ethereum ecosystem. Following the hack, the community discussed two options on how to deal with it: i) do nothing, following the rule of code is law and the immutability of the blockchain, and ii) replay the hack and pay out the funds to the original contributors. What is known today as Ethereum went for option 2), whereas a smaller part of the community disagreed and went with option 1) which is known nowadays as Ethereum Classic. Forking to some extent can be compared to voting by the feet as observed in many democracies such as Switzerland where people in disagreement with a policy change in a district move to another one being closer to their values and perspectives \cite{somin2020free,liebig2007taxation}. The study of the observed forks in DAOs could inform insights on the governing mechanisms of this voting by the feet.
The fork and the replay of the blockchain also demonstrate that blockchain systems are in fact \textit{socio}-techno-economic systems whose behavior cannot be purely derived from the technical configuration of the underlying technology stack. In particular, properties such as immutability are emergent within blockchain systems and depend on the community maintaining and contributing to a particular blockchain system. In this sense, two communities can maintain the same technology stack with the same parametrization, but one being more immutable than the other (in the example above, Ethereum Classic being more immutable than Ethereum). 
In particular, mechanisms such as "forking" provide democratic checks, allowing participants to dissent actively without systemic collapse.
The permissionless nature and the transparency of the network can incentivize intrinsically motivated people to contribute to a network, exemplified by Bitcoin in the early days and Bitcoin Lightning node operators nowadays \cite{beres2021cryptoeconomic,gotham2023irrational}: Operating a Bitcoin Lightning node is economically not feasible, but contributors are still doing it. Considering that intrinsic motivation is a better predictor for endurance and creativity in individuals \cite{ballandies2022incentivize}, permissionlessness might be an important design mechanism to build lasting and resilient systems. Its integration in current digital democracy practices has to be studied in greater detail.

DAOs employ diverse voting methods \cite{spychiger2025decision} including quadratic voting, staking, futarchy, conviction voting, and token-based voting, each presenting unique benefits and trade-offs concerning fairness, representation, and efficiency, and thus provide an interesting testbed for digital democracy research. In the following, a brief introduction to each mechanism is given:
\textit{Quadratic voting} enables participants to indicate the intensity of their preferences by allocating multiple votes to proposals of greater importance. Each additional vote on a given issue incurs a cost equal to the square of the total votes cast for that issue, thereby discouraging disproportionate vote aggregation.
This method is particularly well suited to participatory budgeting and other multi‐option contexts, where stakeholders must balance support across several proposals. A prominent instantiation is \textit{quadratic funding}, in which individuals contribute funds to multiple projects without a marginal increase in cost and a public funder matches each project’s total contributions according to \( \bigl(\sum_i \sqrt{c_i}\bigr)^2 \),
where \(c_i\) denotes the amount contributed by donor \(i\). This formula amplifies smaller donations while constraining the influence of large donors.
\textit{Futarchy} is increasingly explored in DAOs \cite{ballandies2024daos}, which was introduced before the concept of DAOs emerged \cite{hanson2013shall}. It is a prediction-market-based approach that leverages collective forecasting to select policies likely to yield the best measurable outcomes, thereby incentivizing accuracy and effectiveness rather than popularity. \textit{Conviction voting} facilitates continuous voting on proposals with a vote's power given for a proposal increasing over time. A proposal is then accepted once a threshold is reached, enabling minority groups with consistent interests to gradually accumulate enough support to pass proposals.
\textit{Optimistic governance} assumes that proposals pass by default unless a specified rejection threshold of objections is met, rather than requiring an explicit quorum for approval. This “default‑pass” model can greatly streamline decision‑making and reduce voter fatigue, though it also raises the risk that marginal or contentious proposals receive insufficient scrutiny.
Token-based governance grants voting rights directly based on token ownership, aligning influence with economic stake, though at the risk of governance dominated by wealthier participants. 
\textit{Square‑root voting} adapts the same principle as quadratic voting to token‑based governance. Here, voting power is proportional to the square root of the tokens held, such that each additional token yields diminishing marginal influence. Unlike quadratic funding where donors allocate their funds across projects, square‑root voting does not transfer tokens to a common pool; instead, each participant’s voting power is computed once for a single decision. This approach is generally employed for binary or single‑issue votes (for example, yes/no resolutions in associations \cite{ballandies2023onocoy}), rather than for allocating resources across multiple alternatives.
Both mechanisms require robust participant authentication to prevent Sybil attacks, in which malicious actors create multiple identities to amplify their influence by circumventing the marginal gains calculation of voting power.
\textit{Staking} assigns voting power proportional to the amount of tokens or resources participants have invested or locked into the system, which rewards commitment.
\textit{Hash-power voting} reflects the share of computational work each voter contributes (e.g., in proof-of-Work systems); it aligns incentives with network security but can centralize around industrial-scale operators. 
\textit{Economic-majority} or \textit{user-activated soft forks} utilize the permissionless nature of blockchain systems and rely on the aggregate real-world economic power (e.g. fiat currencies, but also goods and services) of nodes, exchanges, and users who upgrade software and collectively reject non-compliant decisions from consensus node operators, enabling grassroots veto power over those having a larger power in the consensus of stake-based or hash-power-based voting mechanisms, but at the cost of complex coordination and other challenges such as manipulation from those owning larger amounts of fiat currencies (see Section \ref{sec:risks}). 
\textit{Time-weighted voting} \cite{wang2025balancing} assigns greater voting power to token units that are held for a larger period of time, thus rewards long-term holders of a token with greater voting power, amongst others mitigating flash-loan attacks on voting (see Section \ref{sec:risks}).

Besides voting, tokens in DAOs can facilitate the adaptation of their complex system via improved real-time feedback \cite{dapp2021finance,kleineberg2021social,ballandies2022incentivize,spychiger2025short}. Tokens are given to DAO participants for the decentralized execution of tasks. For instance, Bitcoin directly utilizes tokens to incentivize the desirable behavior of contributing to its consensus mechanism. Also, bounty rewards or other desirable actions such as the placement of infrastructure nodes \cite{ballandies2023taxonomy}, joining the decision-making processes, or providing liquidity can be incentivized by DAOs. Further, the token price can be seen as a further signal to the community on their overall performance \cite{ballandies2025bitcoin}. In general, bottom-up created token incentives can be a suitable tool for democracies to facilitate sustainable actions efficiently \cite{dapp2021finance,ballandies2021finance}, and thus be an alternative to centralized and hierarchical control approaches in steering societies. Exploring the use of token incentives in digital democracy environments could thus, in theory, facilitate more efficient democracies. However, this needs to be carefully studied and implemented, as the intrinsic motivation of participants might crowd out or, in general, the quantification of behaviors might not align with the values of system stakeholders which would result in their potential rejection of the implemented system \cite{ballandies2022incentivize}.  

\subsection{Improving DAOs}
\label{sec:improving_daos}
Drawing on research in digital democracy, DAO designs can benefit from existing theoretical and empirical insights.
Digital democracy consists of two main components, deliberation and voting \cite{helbing2023democracy}. Although voting is extensively utilized and innovated within DAOs (Section \ref{sec:dao_innovations}), deliberation is an overlooked component \cite{ballandies2025bitcoin}. Usually, DAOs utilize the deliberation mechanism inspired by Bitcoin, which is referred to as the Improvement Proposal mechanism \cite{ballandies2025bitcoin}: a contributor drafts a markdown document that explains the motivation, precise specification, and backwards-compatibility plan; submits it publicly, usually as a GitHub pull request; receives a sequential number and category from editors; then iterates through an open-comment period where different stakeholders critique the design until rough community consensus is reached. Once deemed “Final” by the proposal initiator, it becomes subject to voting. Because these documents are version-controlled, every decision is permanently recorded, giving the ecosystem a shared technical memory while ensuring that only broadly supported ideas become part of the deployed protocol.
Nevertheless, this proposal mechanism is a very rudimentary process with several limitations ranging from challenges in having broad parts of the community participate, over high barriers for the broader community to formulate a well-designed proposal because of their academic and technical nature, and the long time it takes to draft and decide on a proposal.
Recent work in digital democracy \cite{yang2025bridging} explored three deliberation mechanisms in the context of participatory budgeting that could be explored within DAOs: 
(i) Preference-based Clustering maps approval votes, slices the map into wedges, and places like- or unlike-minded people in balanced groups for discussions; 
(ii) a slider-controlled, human-in-the-loop Method of Equal Shares previews how much of the budget an algorithm would allocate before a public discussion; 
(iii) ReadTheRoom projects survey statements, lets participants place them along a spectrum, then allows opinion shifts after discussion.  
In Swiss and Taiwanese pilots, the toolkit balanced groups and halved the polarization. Although originally implemented requiring physical presence, these mechanisms might be generalizable to purely online audiences in DAOs.

Voter turnout in DAOs is frequently low and has drawn criticism, and DAO literature increasingly suggests that most users are simply uninterested in governance \cite{liu2024illusion,meneguzzo2025evaluating}. A similar phenomenon appears in direct‑democracy votes: participation is often low and declining, leading many to doubt the efficacy of direct democracy. However, Swiss referendum studies highlight cumulative engagement as a more meaningful indicator than isolated turnout rates \cite{serdult2021referendum,serdult2013partizipation}. Over a four‑year period, 80–90\% of the population participates in at least one referendum, versus an average turnout of 45 \% at any single vote \cite{serdult2021referendum}. A comparable pattern may exist in DAOs and warrants further investigation. Moreover, digital voting has not been shown to boost turnout \cite{germann2017internet}, so DAO designers should not assume that more efficient voting tools alone will raise participation.

Experimental evidence suggests that participants’ sense of procedural legitimacy in digital participatory budgeting is shaped by two distinct design choices: the expressiveness of the ballot interface and the distributive logic of the aggregation rule. In a controlled lab study \cite{yang2024designing}, it has been found that moderately expressive ballots, such as rank-ordered or point-allocation formats, balance cognitive effort with preference articulation; yet aggregation exerts the stronger influence on procedural justice: the proportional Method of Equal Shares (ES) yields more egalitarian outcomes and higher fairness and trust ratings than the common Greedy heuristic, provided that participants receive a brief explanation of the rule. Complementary field data~\cite{hausladen2024voting} show that legitimacy assessments of majority, approval, range, and modified Borda voting shift with issue salience and preference clarity: simple majority fares best in low-stakes, non-polarised settings, whereas range and preferential rules are preferred for high-stakes, polarised decisions, particularly among voters with well-defined preferences. Together, the studies imply that DAO and other digital governance systems should treat ballot design and aggregation as an integrated user experience, calibrating expressive capacity and adopting transparent, proportional algorithms such as ES, while retaining flexibility to adjust these parameters as issue characteristics and electorate competence evolve \cite{hausladen2024voting,yang2024designing}.

Also the verifiability of digital voting channels is important as it can raise the trust in digital voting. For instance on-paper return codes, public test site, and external security audits have been found to increases trust into digital voting \cite{milic2016haltungen} and might improve the User experience and trust in DAOs.
Moreover, campaign financing of elections/ referendums are an important part of demcoracy \cite{dawood2015campaign} and an active research field. For instance, the city of Seattle tested campaign vouchers \cite{mccabe2019diversifying} to add transparency and fairness to the financing of elections. However, campaign financing in DAOs has not been studied so far. In general, DAOs could be an interesting testbed to analyze campaign financing and experiment with novel solutions.

\section{DAOs for Democratic Human-Machine Collaboration}
\label{sec:dao_human_machine}

Artificial intelligence (AI) can augment DAOs in multiple, mutually reinforcing ways. First, AI embedded in DAO tooling can summarise complex governance proposals, optimise treasury management, and even execute delegate voting while preserving the preferences of token holders. Second, the community-owned nature of DAOs provides a credible path towards building resilient, trustworthy, and efficient decentralised infrastructures on which AI services can run. Third, because DAOs rely on transparent, permissionless, and collectively verifiable rules, they can facilitate a principled framework for integrating such AI capabilities without eroding human autonomy or accountability.
Taken together, these features can pave the way for a symbiotic human–machine collaborations that facilitates and enhances self-organization.

This section develops this argument in three steps. Section \ref{sec:ai_in_dao} (“AI in DAOs”) analyses the concrete roles AI can play inside DAO governance and execution processes. Section \ref{sec:depai} (“Decentralised Physical AI”) shows how DAOs can serve as a democratic organisational structure for decentralized energy, data, compute and model-sharing infrastructure networks. Section \ref{sec:collective_human_agent} (“Human–Agent Collaboration”) explores self-organization that emerges when autonomous agents, robots and human contributors co-create knowledge, make decisions and execute them under DAO-mediated incentives.

\subsection{AI in DAOs}
\label{sec:ai_in_dao}

AI utilized in DAOs can enhance both, their efficiency and security. In particular, a growing stream of work argues that AI improves coordination and adaptability in decentralized organizations \cite{mcconaghy2016aidaos, wang2019decentralized, aragon2023aidaos, buterin2024cryptoai}.

Large language models (LLMs) can for instance streamline governance by classifying proposals, thus making assessment more efficient for community members \cite{ziegler2024classifying}. Furthermore, LLM pipelines utilizing chain-of-thought reasoning can summarise proposals and generate stakeholder-specific voting recommendations, potentially improving voter turnout by removing barriers in understanding and judging proposals \cite{chen2025intelligent}. Another AI-driven approach employs natural language processing (NLP) to estimate proposal passage likelihood under full participation, allowing strategic voting with sampled subsets of members, thus potentially enhancing decision-making efficiency by involving smaller but representative participant groups \cite{ashkenazy2023efficient}.
In parallel, DAOs are beginning to route execution to specialized agent systems: Web3-native autonomous agents capable of blockchain interactions can facilitate efficient management of treasury funds and the tokenization of real-world assets and infrastructures, reducing operational overhead and increasing responsiveness \cite{ballandies2023taxonomy,walters2025eliza, borjigin2025ai}. For example, SingularityDAO and ElizaOS deploy AI agents to execute trades automatically in portfolio optimizations, and Virtuals allows users to govern tokenized AI agents, directing their design and profit generation. 
Furthermore, AI-powered agents can draft and commit governance proposals on-chain, significantly reducing proposal preparation time  \cite{ao2025agentdao}. They can also vote on behalf of humans, such as in the UOMI Network, where agents cast votes for tokenholders according to predefined on-chain rules. Together, these approaches streamline decision-making processes. 

AI systems can also detect governance risks, such as proposals whose on-chain implementation diverges significantly from their off-chain descriptions, thereby flagging potentially malicious intentions \cite{ma2024demystifying}. Similarly, Graph Convolutional Neural Network-based autoencoders can detect and eliminate Sybil identities from DAO voting networks, improving resistance to manipulation \cite{dupont2023new}.

Collectively, these AI-driven enhancements within DAOs can facilitate more efficient, resilient and secure governance processes.

\subsection{Decentralized physical AI via DePIN}
\label{sec:depai}

DAOs are also an enabler of decentralized physical infrastructure networks (DePIN) \cite{ballandies2023taxonomy}. These are systems that utilize cryptoeconomics (Section \ref{sec:cryptoeconomics}) to build infrastructures such as telecommunication, weather stations, or compute networks. For these systems to be decentralized, the governance is also required to be decentralized~\cite{ballandies2023taxonomy}. Hence, these systems often utilize DAO communities to place and maintain hardware devices for which these operators receive token rewards in return. The token is then used by 3rd parties to access system services \cite{kalabic2023burn} and by the community to participate in governance, both giving value to the token and closing the loop in the local economy.

\begin{figure}
    \centering
     \caption{The DePAI Stack: A Vertically Integrated Architecture for Decentralized Physical Artificial Intelligence.}
    \label{fig:depai}
    \includegraphics[width=0.6\linewidth]{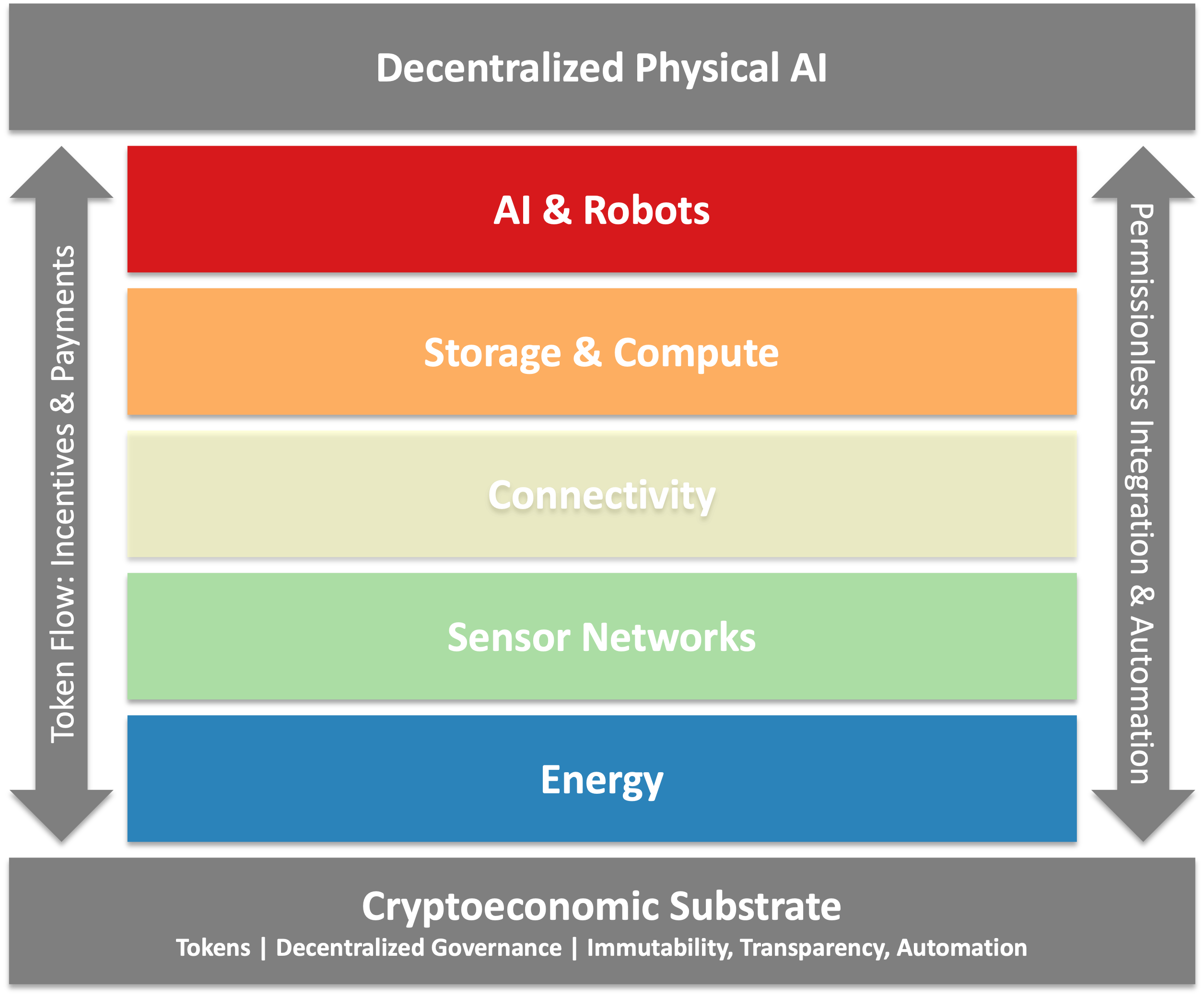}
\end{figure}

Figure \ref{fig:depai} positions DePIN systems along a vertically integrated stack. At the foundation, decentralized energy networks power the entire infrastructure, enabling autonomous, resilient operations independent of centralized utilities. Building on this foundation, decentralized sensor grids capture real-world signals such as weather, RF, and GNSS corrections, and relay them through permissionless connectivity protocols. For instance, in onocoy \cite{ballandies2023onocoy}, users, in return for token incentives, autonomously place GNSS atennas on their properties to share real-time kinematic correction data with the network, which uses this information to improve GPS positioning for their customers. 
Such data enters decentralized storage networks (e.g. Filecoin \cite{guidi2022evaluating}), flows into distributed compute (e.g. golem \cite{kondru2020review}), and ultimately supports AI workloads, ranging from autonomous agents to the control of robotic systems. Each layer is independently operated yet economically linked via tokens, shared governance and the underlying blockchain, forming what is termed decentralized physical AI (DePAI): a full-stack, trust-minimized substrate where every byte, packet, and model weight is collectively owned and auditable by decentralized communities. Merging these verticals can result in self-governing entities, as pioneered with the concept of a self-owning house \cite{hunhevicz2021no1s1}, which has subsequently been further studied~\cite{wang2025automation,spychiger2024governance,lustenberger2025daos}. Also the concept of robot swarms \cite{strobel2023robot,pacheco2020blockchain,pacheco2025swarm} is noteworthy in this context. These are autonomously AI-controlled robots where the intelligence lies within a robot and that can interact in peer-to-peer interactions with others. On the one hand, these networks can be understood as a sensing network within the DePAI Stack by enabling robots be send task specifically to a particular physcial location. On the other hand, these robot swarms form a decentralized physical network themselves. Lastly, they enable the execution of digital decisions in the physical world, by letting robots enact those decisions.

Because a single centralized layer would re-introduce a point of failure, DAO principles must pervade the entire stack, not just the AI-application tier. When sensors, bandwidth, storage, and compute are provisioned and steered by token-aligned communities, the result is a tamper-resistant, co-created infrastructure with built-in checks and balances. Such DAO-governed DePAI systems can mitigate manipulation risk, increase resilience, and invite democratic oversight principles. In particular, when AI applications are built on such decentralized foundations, they may become robust platforms for human-machine self-organization.

\subsection{Human-machine collaboration for enhanced self-organization}
\label{sec:collective_human_agent}

\begin{figure}
    \centering
     \caption{Human–machine collaboration for self-organization. Local human and machine agents autonomously (i) decide how to respond to changing environments and (ii) share in the governance that defines their interactions. A permissionless architecture—foundational to human autonomy—underpins the emergence of self-organization.
}
    \label{fig:human_machine_collab}
    \includegraphics[width=0.7\linewidth]{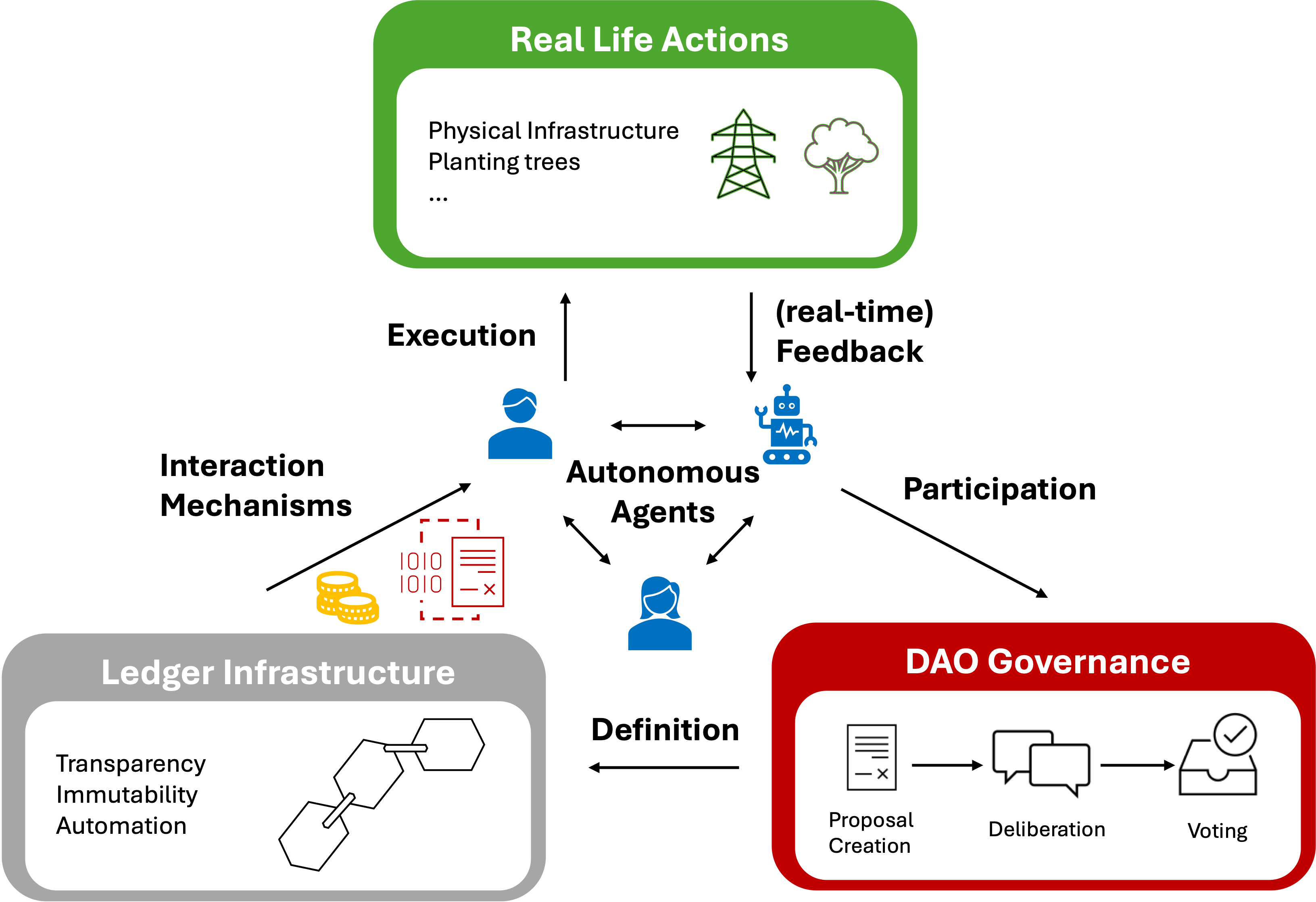}
   
\end{figure}

AI agents increasingly automate task execution for humans (Section \ref{sec:ai_in_dao}), and decentralized autonomous organizations provide a coherent framework for governing these interactions \cite{chaffer2024decentralized}. 
In particular, the value of transparency within DAOs if implemented well can clearly delineate the interfaces at which human inputs and AI-generated actions intersect and the rules on how these interactions are characterized. Such a transparent infrastructure could mitigate manipulation risks, thereby fostering trust necessary for sustainable adoption and efficient execution, as illustrated in Figure \ref{fig:human_machine_collab}.

Here, the value of permissionlessness that DAOs introduce to digital democracy can, if implemented correctly, safeguard the autonomy of human actors which is necessary for self-organization to emerge \cite{ballandies2024daos}. Value-sensitive design methodologies \cite{ballandies2024advancing,mahajan2025co} have to be applied here to facilitate the incorporation of such values, eventually empowering participants to engage voluntarily, retain control over their contributions, and choose and meaningfully exit their memberships. At the same time, this autonomy supports adaptation: human actors within a DAO can autonomously determine how to enact collective decisions in response to local conditions and feedback, enabling efficient adjustment to changing environments~\cite{kleineberg2021social,dapp2021finance}, arguably the key success principle of Bitcoin \cite{ballandies2025bitcoin}. Extending this paradigm of local autonomy to software agents, robots, and other technical devices could further enhance system adaptability, as earlier illustrated by decentralized traffic-light coordination \cite{lammer2008self,lammer2016selbst}, or more recently with robot swarms sensing enviornmental states trustworthily \cite{pacheco2025swarm}. In particular, DAOs provide an efficient and resilient framework for the bottom-up organization of these complex human-machine interactions. 

The voting mechanisms explored in DAOs can inform the integration of physical AI agents in human decision-making, such as the optimistic governance mechanisms, where AI agents undertake micro-decisions subject to potential human override. Such a governance framework allows AI efficiency to be balanced with human oversight, maintaining accountability and preventing runaway outcomes that could compromise collective objectives and values. 

Moreover, concepts explored in the context of DAOs such as prediction markets and Futarchy can extend this human-machine collaboration even further: The Netflix Prize famously showed that ensembling multiple optimisation algorithms beats any single model \cite{helbing2021networked}, and complexity research finds that groups of people, under the right incentives, can also out-predict even high-performing individuals \cite{hong2004groups}, an effect which is referred to as collective intelligence \cite{woolley2010evidence}. Bringing these two strands together in one system, where humans and autonomous agents collaborate could facilitate synergies increasing overall forecasting accuracy.
DAOs are a natural vehicle for such a collaboration that maintains democratic oversight. By facilitating interactions transparently on-chain, a DAO can integrate machine forecasts and human intelligence into the same cryptoeconomic fabric, incentivizing those mechanisms and signals which move the system closer to the truth, eventually facilitating collective learning. This decentralised structure could aggregate diverse perspectives, limit single points of failure, and dampen manipulation, ultimately enabling a more trustworthy human-machine collaboration which would let an enhanced collective intelligence emerge.

Finally, through DePIN, DAOs reach beyond the digital realm into the physical world, orchestrating networks of decentralised autonomous machines \cite{castillo2025trustworthy,pacheco2025swarm} and infrastructures. Here, AI would supply the intelligence for automation, DePIN provides real‑world integration, and the DAO facilitates governance \cite{castillo2025trustworthy} under human oversight, together forming a democratic, resilient, and self‑improving socio‑techno-economic system in which both, local human and machine agents autonomously decide how to respond to changing system environments, which is a prerequiste for self-organization to emerge \cite{ballandies2024daos}. 



\section{Challenges and Risks in decentralized physical AI}
\label{sec:risks}
DAOs and DePAI are promising to create democratic human-machine collaborations facilitating self-organization (Section \ref{sec:collective_human_agent}), but they also face notable challenges and risks as explored in the following. 

While AI can enhance DAO coordination and decision support, we explicitly acknowledge risks. LLMs remain vulnerable to attacks, such as prompt injection \cite{greshake2023not,rossi2024early}, transferable jailbreaks \cite{zou2023universal}, and social bias \cite{gallegos2024bias}, which undermine systems that depend on them. The blockchain layer faces risks as well: although interaction rules in DAOs are transparently encoded on-chain via smart contracts, they remain susceptible to security breaches. These vulnerabilities call into question the viability of fully automated solutions and underscore the need for human oversight, including the capacity to “pull the plug.” Such oversight is compatible with the permissionless ethos of DAOs and is necessitated by the autonomy of their contributing agents, yet how to guarantee this principle in practice remains unclear. Also, in general, participants may underestimate the potential for manipulation and hacks in IT systems and machines, treating decisions and executions within DAO-based human-machine collaborations as infallible. For instance, despite the 2016 DAO hack and the lessons it offered (Section \ref{sec:dao_innovations}), the Ethereum Foundation (the entity mainly impacting the development of the Ethereum blockchain) continues to state on its website: “Once the contract is live on Ethereum, no one can change the rules except by a vote. If anyone tries to do something that’s not covered by the rules and logic in the code, it will fail. And because the treasury is defined by the smart contract too, that means \textit{no one can spend the money without the group’s approval} either” \cite{ethereumDAO}. This overlooks the historical fact the Ethereum community itself experienced during the DAO hack when an attacker diverted funds \textit{without} the group’s approval. Accordingly, doubts persist about whether human-machine collaboration can be truly empowering when it ultimately relies on human users maintaining a critical, hands-on relationship with the technology.

A Value-sensitive design approach might facilitate and re-emphasise the need for human integration and oversight in these systems, as it had been performed in the past for blockchain systems \cite{ballandies2024advancing,ietto2023blockchain}. Recent work emphasises the iterative value articulation during design and the continued involvement of humans in shaping the governance and design of a human-AI system, in particular once the system is deployed \cite{mahajan2025co}; these continuous evaluations could be combined with risk evaluation frameworks \cite{ai2023artificial,liang2022holistic}. Considering the DAO hack, this is very much required and a crucial component, but it is still an open question how to facilitate it under the time and financial constraints of real-world system instantiations. More practical research and exploration in this direction are required. 

Furthermore, though voting mechanisms are extensively explored, deliberation is often overlooked in DAOs~\cite{ballandies2024daos,ballandies2025bitcoin}. This limited exploration of deliberation in DAOs increases the risk of having manipulated or biased vote objects on which the community is allowed to vote. Insights from Digital Democracy research should be incorporated in living DAOs (see Section \ref{sec:improving_daos}).  
The concept of the economic majority in Bitcoin-like DAOs in which real-world values, goods and services have the power to decide exhibit the risk of larger monetary stakeholders enforcing protocol changes unilaterally \cite{ballandies2025bitcoin}, thus reintroducing real-world imbalances into DAOs. 

Moreover, the use of token incentives in the context of voting might bias decisions and potentially crowd out intrinsic motivation to participate in those votes \cite{ballandies2022incentivize}. In particular, economic incentives within DAOs can steer actors to game the system, such as utilizing flash-loans to quickly obtain voting power in large amounts and then to approve and execute malicious governance proposals  \cite{wang2025balancing}.  In general, poorly designed incentives can trigger undesirable emergent behaviors \cite{spychiger2024governance} such as "pump-and-dump" schemes \cite{bolz2024machine}, often driven by extrinsically motivated individuals. Also they can reduce the decentralization in a network, which hinders the emergence of adaptation and collective intelligence in DAOs \cite{ballandies2025bitcoin}. For instance, incentive structures emphasizing agent centrality typically foster core-periphery network architectures \cite{konig2014nestedness,konig2010assortative}. Such centralization is observed widely across blockchain ecosystems:
Empirical evidence highlights centralization trends regarding wealth concentration \cite{kondor2014rich,di2018data,de2021heterogeneous}, microvelocity of transactions \cite{de2025microvelocity}, peer-to-peer network topology \cite{gao2025monero}, and within decentralized finance platforms \cite{yan2025network,eisermann2025concentration}.

In addition, legal challenges exist. Legal structures might not resemble the presented realities and impacts of voting participation. For instance, in the Helium network, one of the first DePINs, votes on network matters are continuously performed. However, these votes seem not to be binding for the legal entity behind Helium. In particular, they are only facultative as it was demonstrated with the vote on Helium improvement proposal nr. 138. The necessary required turnout for accepting the proposal was not reached. However, Helium's management approached a decisive voter counter to the proposal off-chain and made him state publicly that he changed his vote to acceptance. Helium then accepted the vote without repeating it on-chain
\footnote{\url{https://www.reddit.com/r/HeliumNetwork/comments/1h1fos9/update_on_hip_138/}, last accessed: 2025-07-25}.
This illustrates that the votes performed on-chain are non-binding. A potential DAO-enabled physical AI system would require making explicit and transparent how decisions are actually taken and what their legally binding character is, so that participants can meaningfully reflect on the decentralized nature of the system in which they participate beforehand, mitigating a potential tyranny of structurelessness \cite{freeman1972tyranny}.

In conclusion, while DAOs offer significant promise for facilitating democratic human-machine collaboration and the self-organization of this integration, they must navigate substantial challenges related to centralization, incentive misalignments, real-world power structures, and technical challenges. Addressing them systematically will be important for fully realizing the potential of the self-organization instantiated by DAOs.

\section{Conclusion}
\label{sec:conclusion}

DAOs represent a promising paradigm for an ethical and democratic human-machine collaboration, advancing self-organization of human society through decentralized governance and cryptoeconomic incentives. 
In particular, they can exemplify human-machine collaboration by embedding AI-driven processes  and autonomous robots within democratic frameworks. Through transparent and permissionless designs, DAOs might foster environments where AI agents perform micro-decisions subject to human oversight, creating an "optimistic governance" paradigm that balances efficiency with autonomy. Also, DAOs can be the enabler of a decentralized data collection, storage and compute infrastructure that facilitates a democratic human-machine system in the first place. Finally, by integrating autonomous robots in this decentralized framework, a resilient and efficient bottom-up collaboration might emerge that self-organizes in response to local environments and feedback.

However, realizing the full potential of this collaborative framework requires addressing centralization risks in cryptoeconomic networks, designing effective incentives, and applying insights from digital democracy research systematically into DAOs. By navigating these tasks, DAOs can genuinely embody democratic participation, collective intelligence and efficient adaptation in the age of AI and robots. The challenges and risks illustrated in this chapter emphasize the necessity for careful design and continued assessment and refinement of such systems, in particular, after their deployment. A DAO-based human-machine system should not be considered as final or infallible, but as an instantiation that is subject to changes according to the values of its participating members and emerging findings from research and practice. 
How to facilitate this effectively and resiliently remains an open question.


\clearpage
\bibliographystyle{apalike}
\bibliography{sample}

\end{document}